\documentclass[11pt]{amsart}
\usepackage{amsmath, amssymb, amsthm, mathtools, abraces, xcolor}
\usepackage{relsize, paralist, hyperref, cite, setspace, enumitem}
\usepackage[paper=a4paper,left=3cm,right=3cm,top=3cm,bottom=3cm]{geometry}

\newtheorem{theorem}{Theorem}[section]
\newtheorem{proposition}[theorem]{Proposition}

\newtheorem{lemma}[theorem]{Lemma}

  \theoremstyle{definition}

\newtheorem{example}[theorem]{Example}

  \theoremstyle{remark}
\newtheorem{remark}[theorem]{Remark}

\numberwithin{equation}{section}

\newcommand{ \C }{ \mathcal C }
\newcommand{ \A }{ \mathcal A }
\newcommand{ \myx }{ \boldsymbol{x} }
\newcommand{ \myy }{ \boldsymbol{y} }
\newcommand{ \myz }{ \boldsymbol{z} }
\newcommand{ \Irr }{ \operatorname{Irr} }
\newcommand{ \Prob }{ \operatorname{Pr} }
\newcommand{ \leqthree }{ {\leqslant\!3} }
\newcommand{ \leql }{ {\leqslant\!\ell} }
\newcommand{ \bs }[1]{ \boldsymbol{#1} }
\newcommand{ \sml }{ \mathsmaller }

\newcommand{ \defeq }{ \coloneqq }
\newcommand{ \myqed }{ \hfill $\blacktriangle$ }

\begin{document}

\title[Codes correcting short tandem duplications]
			{On the Maximum Number of Non-Confusable Strings Evolving Under Short Tandem Duplications}

\author{Mladen~Kova\v{c}evi\'{c}}

\thanks{The author is with the Faculty of Technical Sciences, University of Novi Sad, 21000 Novi Sad, Serbia.
Email: kmladen@uns.ac.rs. ORCiD: \href{https://orcid.org/0000-0002-2395-7628}{0000-0002-2395-7628}.}

\thanks{This work was supported by the European Union's Horizon 2020
research and innovation programme under Grant Agreement no. 856967,
and by the Secretariat for Higher Education and Scientific Research
of the Autonomous Province of Vojvodina through the project no. 142-451-2686/2021.}

\subjclass[2020]{94A24, 94A40, 94B25, 94B50, 68R15.}

\date{April 24, 2022.}

\keywords{Tandem duplication, tandem repeat, duplication error, repetition
error, sticky insertion, DNA storage, error correction, zero-error capacity,
constrained code, square-free string.}

\begin{abstract}
The set of all $ q $-ary strings that do not contain repeated substrings
of length $ {\leqslant\! 3} $ (i.e., that do not contain substrings of
the form $ a a $, $ a b a b $, and $ a b c a b c $) constitutes a code
correcting an arbitrary number of tandem-duplication mutations of length
$ {\leqslant\! 3} $.
In other words, any two such strings are non-confusable in the sense that
they cannot produce the same string while evolving under tandem duplications
of length $ {\leqslant\! 3} $.
We demonstrate that this code is asymptotically optimal in terms of rate,
meaning that it represents the largest set of non-confusable strings up to
subexponential factors.
This result settles the zero-error capacity problem for the last remaining
case of tandem-duplication channels satisfying the ``root-uniqueness''
property.
\end{abstract}

\maketitle

\section{Introduction}
\label{sec:intro}

Tandem duplications are a type of ``sticky'' errors that naturally occur as
mutations in DNA strings and are therefore a potential source of impairments
in \emph{in vivo} DNA-based data storage systems \cite{jain2}.
While the problem of correcting tandem duplications of fixed and known length
$ \ell $ is well-understood, both in scenarios with bounded
\cite{kovacevic+tan, lenz} and unbounded number of errors \cite{jain2, kovacevic},
much less is known about the presumably more relevant problem of correcting
duplications of varying lengths.
For example, optimal codes correcting all patterns of duplications of length
$ {\leql} $ have been found only in the special cases $ \ell = 1 $ and
$ \ell = 2 $ \cite[Thm~32]{jain2}.
Our main contribution here is a proof that an analogous construction of codes
correcting an unbounded number of tandem duplications of length $ \leqthree $
\cite[Thm~27]{jain2} is also asymptotically optimal in terms of rate.
This result settles the zero-error capacity problem for tandem-duplication
channels in all cases where the duplication roots of strings are unique
\cite[Thm~40]{jain2}.
For larger values of $ \ell $, however, the roots with respect to duplications
of length $ \leql $ are not unique \cite[Thm~40]{jain2} and, hence, different
constructions and upper bounds will be required to solve the zero-error
capacity and related problems for these models.%
\footnote{The first code constructions for these models (with
$ \ell \in \{4, 5, \ldots \} $) have been reported in \cite{chee2}.}

Apart from information- and coding-theoretic questions of the kind we
discuss here, several other problems concerning models with tandem
duplications of varying lengths have been studied in the literature;
see, e.g., \cite{farnoud, jain, leupold}.

\pagebreak
\subsection{Model description}
\label{sec:channel}

The $ q $-ary alphabet is denoted by $ \A_q \defeq \{0, 1, \ldots, q-1\} $, and
the set of all strings (or words) over $ \A_q $ by $ \A_q^* \defeq \bigcup_{n=0}^\infty \A_q^n $.
The length of a string $ \myx = x_1 \cdots x_n \in \A_q^n $ is denoted by $ |\myx| = n $.
The string obtained by concatenating two strings $ \bs{u} $ and $ \bs{v} $ is
written as $ \bs{u} \bs{v} $.
A string $ \bs{v} $ is said to be a substring of $ \bs{x} $ (or a segment in
$ \bs{x} $) if there exist (possibly empty) strings $ \bs{u} $ and $ \bs{w} $
such that $ \bs{x} = \bs{u} \bs{v} \bs{w} $.

Let $ \ell $ be a fixed positive integer.
The \emph{$ (\leql) $-tandem-duplication channel} acts on a transmitted
string $ \myx $ by successively applying to it a number of tandem duplications,
each of length $ \leql $, where a tandem duplication of length $ k $ is
an insertion of an exact copy of a substring of length $ k $ next to the
original substring (it is irrelevant whether a duplicate is inserted to
the left or to the right of the original as both options result in the
same string).
We assume that the number of applied duplications is not known in advance to
either the transmitter or the receiver and can take on any value in the set
of natural numbers $ \{0, 1, 2, \ldots\} $.
In more precise terms, the channel is described as follows:
\begin{itemize}[leftmargin=0.6cm]
\item
Input: $ \myx \equiv \myx^\sml{(0)} $
\item
Choose a number $ t $ (the number of duplications) from the set
$ \{0, 1, 2, \ldots\} $
\item
For $ i = 1, \ldots, t $, repeat the following:
\begin{itemize}
\item
Choose the duplication location $ j $ in the string $ \myx^\sml{(i-1)} $
arbitrarily from the set $ \{1, \ldots, |\myx^\sml{(i-1)}|\} $
\item
Choose the duplication length $ k $ arbitrarily from the set $ \{1, \ldots, \min\{j, \ell\} \} $
\item
Insert a copy of the substring $ x_{j-k+1}^\sml{(i-1)} \cdots x_j^\sml{(i-1)} $
next to the original substring in $ \myx^\sml{(i-1)} $ to produce $ \myx^\sml{(i)} $,
that is
\[ \phantom{XXXXX}
   \myx^\sml{(i)} = x_1^\sml{(i-1)} \cdots \overline{ x_{j-k+1}^\sml{(i-1)} \cdots x_j^\sml{(i-1)} }
   \underline{ x_{j-k+1}^\sml{(i-1)} \cdots x_j^\sml{(i-1)}} x_{j+1}^\sml{(i-1)} \cdots x_{|\myx^\sml{(i-1)}|}^\sml{(i-1)}  \ , \]
where the original substring that is being duplicated is overlined, and
the inserted duplicate is underlined
\end{itemize}
\item
Output: $ \myy \equiv \myx^\sml{(t)} $.
\end{itemize}
Hereafter we assume that $ \ell = 3 $.
\begin{example}
The following list of strings, each producing the next via a tandem
duplication of length $ \leqthree $, is an example of how the channel
acts on a transmitted string $ \myx \in \A_3^8 $:%
\begin{subequations}
\label{eq:desc0}
\begin{alignat}{2}
\label{eq:desc01}
  &\myx             &&=  \ 0 \ 1 \ 1 \ 2 \ 0\ 2 \ 1 \ 0   \\
  &\myx^\sml{(1)}   &&=  \ \overline{0} \ \underline{0} \ 1 \ 1 \ 2 \ 0\ 2 \ 1 \ 0   \\
	&\myx^\sml{(2)}   &&=  \ 0 \ 0 \ 1 \ 1 \ \overline{ 2 \ 0 \ 2 } \ \underline{ 2 \ 0 \ 2 } \ 1 \ 0  \\
	&\myx^\sml{(3)}   &&=  \ 0 \ 0 \ 1 \ 1 \ 2 \ 0 \ 2 \ 2 \ 0 \ \overline{ 2 \ 1} \ \underline{ 2 \ 1 } \ 0  \\
  &\myx^\sml{(4)}   &&=  \ 0 \ 0 \ \overline{ 1 \ 1 } \ \underline{ 1 \ 1 } \ 2 \ 0 \ 2 \ 2 \ 0 \ 2 \ 1 \ 2 \ 1 \ 0  \ .
\end{alignat}
\end{subequations}
Here $ t = 4 $ and the channel output is $ \myy = \myx^\sml{(4)} $.
\myqed
\end{example}

We say that a string $ \myy $ is a $ t $-descendant of $ \myx $, or that
$ \myx $ is a $ t $-ancestor of $ \myy $, if $ \myy $ can be obtained by
successively applying $ t $ tandem duplications of length $ \leqthree $
on $ \myx $.
The set of all $ t $-descendants of $ \myx $ is denoted $ D^t(\myx) $.
Note that a string may belong to both $ D^t(\myx) $ and $ D^s(\myx) $,
$ s \neq t $, because duplications of different lengths are allowed in
the model, i.e., $ D^t(\myx) \cap D^s(\myx) $ is not necessarily empty
(for example, $ 0\,1\,1\,1\,1 $ is both a $ 1 $-descendant of $ 0\,1\,1 $
obtained via a single duplication of length $ 2 $, and a $ 2 $-descendant
of $ 0\,1\,1 $ obtained via two duplications of length $ 1 $ each).
The set of all descendants of $ \myx $ is denoted
$ D^*(\myx) \defeq \bigcup_{t \geqslant 0} D^t(\myx) $, where $ D^0(\myx) \defeq \{\myx\} $.
In this notation, for a given input string $ \myx $, $ D^*(\myx) $ is
the set of possible outputs of the $ (\leqthree) $-tandem-duplication
channel.

\pagebreak
\subsection{Non-confusable strings and error-free communication}
\label{sec:terminology}

Two strings $ \myx, \myy \in \A_q^* $ are said to be confusable in a given
communication channel if they can produce the same string at the output of
that channel; they are said to be non-confusable otherwise.
In our terminology, $ \myx $ and $ \myy $ are confusable if they have a common
descendant, i.e., if $ D^*(\myx) \cap D^*(\myy) \neq \emptyset $.
A set of strings $ \C \subseteq \A_q^* $ is said to be a \emph{zero-error}
code \cite{shannon} for a given channel if every two different codewords
$ \myx, \myy \in \C $ are non-confusable.
Note that a zero-error code is able to correct \emph{all} error patterns
that can be realized in the channel;
namely, any given output string can be unambiguously decoded by the receiver
as there is only one codeword from $ \C $ that could have produced it.
A zero-error code $ \C \subseteq \A_q^n $ is said to be optimal if there is
no other zero-error code $ \C' \subseteq \A_q^n $ such that $ |\C'| > |\C| $.

The rate of a code $ \C \subseteq \A_q^n $, expressed in bits per symbol,
is the exponent $ \frac{1}{n} \log_2 \left| \C \right| $.
The zero-error capacity of a channel with input alphabet $ \A_q $ is the
$ \limsup_{n\to\infty} $ of the rates of optimal zero-error codes in $ \A_q^n $.
This quantity represents the largest number of bits per symbol that can be
transmitted through the given channel in an error-free manner.

\section{Duplication roots and irreducible strings}
\label{sec:roots}

By successively applying the operation of de-duplication, i.e., removing
duplicate substrings of length $ \leqthree $, every string $ \myx $ can
be reduced to its \emph{root} string $ R(\myx) $ which contains no repeated
substrings of length $ \leqthree $.
Furthermore, as shown in \cite[Thm 24]{jain2}, the roots are unique, meaning
that one is guaranteed to end up with the same string regardless of the order
in which de-duplication is performed.
(We emphasize that this ``root uniqueness property'' holds only in models
with tandem duplications of length
\begin{inparaenum}
\item[(i)]
$ =\!\ell $,
\item[(ii)]
$ {\leqslant\!2} $, or
\item[(iii)]
$ \leqthree $.
\end{inparaenum}
It does not hold, for example, in models with tandem duplications of length
$ {\leqslant\!\ell} $, when $ \ell \in \{ 4, 5, \ldots \} $; see \cite[Thm 40]{jain2}.)

In this context, a string that contains no repeated substrings of length
$ \leqthree $ is called \emph{irreducible}.
In other words, a string is irreducible%
\footnote{Irreducible strings are an instance of pattern-avoiding strings,
or constrained strings \cite{marcus}, the set of forbidden patterns being
$ \{a\,a , a\,b\,a\,b , a\,b\,c\,a\,b\,c : a, b, c \in \A_q\} $.}
if it contains no substring of
the form $ a\,a $, $ a\,b\,a\,b $, and $ a\,b\,c\,a\,b\,c $, where
$ a, b, c \in \A_q $.
Let $ \Irr_q $ denote the set of all irreducible strings over $ \A_q $,
$ \Irr_q(n) $ the set of all irreducible strings of length $ n $, and
$ I_q(n) $ the cardinality of the latter, $ I_q(n) \defeq |\Irr_q(n)| $.
It follows from the ``root-uniqueness property'' mentioned above that every
two different irreducible strings $ \myx, \myy \in \Irr_q $ are non-confusable
in the $ (\leqthree) $-tandem-duplication channel, i.e.,
$ D^*(\myx) \cap D^*(\myy) = \emptyset $, and therefore the set $ \Irr_q(n) $
is a zero-error code for this channel \cite[Thm 27]{jain2}.

\underline{In the remainder of the article we assume that $ q \geqslant 3 $}
because the problems we address are trivial when the alphabet is binary.
For example, there are only finitely many irreducible strings over a binary
alphabet, $ \Irr_2 = \{ 0, 1, 01, 10, 010, 101 \} $, and the zero-error
capacity of the $ (\leqthree) $-tandem-duplication channel with binary alphabet
is equal to zero.

Of interest to us here is the asymptotic behavior of the quantity $ I_q(n) $
as $ n \to \infty $, particularly its exponential growth-rate:
\begin{equation}
\label{eq:expirr}
  \iota_q  \defeq  \lim_{n \to \infty} \frac{1}{n} \log_2 I_q(n) .
\end{equation}
The exponent $ \iota_q $ can be characterized by using standard methods
from the theory of constrained systems \cite{marcus}, e.g., as the logarithm
of the largest eigenvalue of the adjacency matrix of a directed graph that
represents the state-diagram of the system generating the irreducible strings.
We shall use here a simpler characterization from \cite[Prop.~2]{chee2}
where it was shown that $ I_q(n) $ satisfies the recurrence relation
$ I_q(n) = (q-2) I_q(n-1) + (q-3)I_q(n-2) + (q-2)I_q(n-3) $, and that,
consequently,
\begin{subequations}
\label{eq:r}
\begin{equation}
  \iota_q = \log_2 r ,
\end{equation}
where $ r $ is the unique positive real root of the polynomial,
$ x^3 - (q-2)x^2 - (q-3)x - (q-2) $, i.e., $ r $ is defined implicitly by:
\begin{equation}
  r^3 - (q-2) r^2 - (q-3) r - (q-2) = 0 , \qquad r > 0 .
\end{equation}
\end{subequations}

In the following lemma we give another characterization of the exponent
$ \iota_q $ for the ternary alphabet ($ q = 3 $), as well as the consequent
lower bound on $ \iota_q $ for larger alphabets, which will be instrumental
in proving our main result (Theorem~\ref{thm:main}).

\begin{lemma}
\label{thm:entropy}
For every $ q \geqslant 3 $ and $ \beta \in [0,1] $,
\begin{align}
\label{eq:entropy}
  \iota_q  \geqslant  \frac{ H(\beta) }{ 1 + 2 \beta } ,
\end{align}
where $ H(\beta) \defeq -\beta \log_2 \beta - (1-\beta) \log_2(1-\beta) $ is
the binary entropy function.
The equality in \eqref{eq:entropy} is attained if and only if $ q = 3 $ and
$ \beta = \bar{\beta} $, where $ \bar{\beta} $ is the unique positive solution
of the equation $ (1 - x)^3 = x $.
\end{lemma}
\begin{proof}
We prove the identity:
\begin{align}
\label{eq:entropy3}
  \iota_3  =  \max_{0 \leqslant \beta \leqslant 1} \frac{ H(\beta) }{ 1 + 2 \beta } ,
\end{align}
from which the statement of the lemma will follow immediately (as $ \iota_q $
is a monotonically increasing function of $ q $).
Equating the derivative of $ \frac{ H(\beta) }{ 1 + 2 \beta } $ to zero,
one finds that the maximizer of this function is the unique positive real
number satisfying the equation $ (1 - x)^3 = x $, call it $ \bar{\beta} $.
The right-hand side of \eqref{eq:entropy3} can then be expressed as:
\begin{align}
\label{eq:max}
  \frac{ H(\bar{\beta}) }{ 1 + 2 \bar{\beta} }
    =  \log_2\!\left( \bar{\beta}^{\frac{-\bar{\beta}}{1 + 2 \bar{\beta}}} \cdot (1 - \bar{\beta})^\frac{-1+\bar{\beta}}{1 + 2 \bar{\beta}} \right) 
    =  - \log_2 \! \left(1 - \bar{\beta}\right) .
\end{align}
On the other hand, we know that $ \iota_3 = \log_2 r $, where $ r $ is
the unique positive real solution of the equation $ x^3 - x^2 - 1 = 0 $
(see \eqref{eq:r}).
Therefore, proving the equality in \eqref{eq:entropy3} is equivalent to proving
that $ - \log_2 (1 - \bar{\beta}) = \log_2 r $, i.e., that $ (1 - \bar{\beta})^{-1} $
is a solution to $ x^3 - x^2 - 1 = 0 $.
This can be verified directly by substituting $ (1 - \bar{\beta})^{-1} $ for $ x $
and using the fact that $ (1 - \bar{\beta})^3 = \bar{\beta} $.
\end{proof}

\section{Confusability of strings in the $ (\leqthree) $-tandem-duplication channel}

In this section we demonstrate several facts about the evolution of strings
under tandem duplications of length $ \leqthree $, the main point of which
is to derive an upper bound on the maximum number of pairwise non-confusable
strings in a given descendant cone $ D^*(\myx) $ (Proposition~\ref{thm:desc}).
For a further study of combinatorial and algorithmic aspects of confusability
in the $ ({\leqslant\!2}) $- and $ (\leqthree) $-tandem-duplication channels,
see \cite{chee}.

The following lemma states that a set of pairwise non-confusable strings,
all of which are $ 1 $-descendants of a given string $ \myx $, can have at
most two elements.
The proof also illustrates the conditions under which two non-confusable
strings may be obtained after applying different mutations on $ \myx $
(see \eqref{eq:dupltypes} ahead).

\begin{lemma}
Consider an arbitrary string $ \myx $, the set of its $ 1 $-descendants
$ D^1(\myx) $, and let $ \C \subseteq D^1(\myx) $ be a zero-error code for the
$ (\leqthree) $-tandem-duplication channel.
Then $ |\C| \leqslant 2 $.
\end{lemma}
\begin{proof}
Consider $ \myx', \myx'' \in D^1(\myx) $, and suppose that the mutations
producing $ \myx' $ and $ \myx'' $ from $ \myx $ are applied on different,
non-overlapping substrings of $ \myx $.
Then $ \myx' $ and $ \myx'' $ are confusable because they have a common
descendant; to see this, perform in $ \myx' $ the duplication that has
produced $ \myx'' $ from $ \myx $, and vice versa.
Now suppose that the duplications producing $ \myx' $ and $ \myx'' $ from
$ \myx $ are applied on overlapping substrings of $ \myx $.
It turns out that in all the possible cases \emph{but one}, we can use the
same reasoning as for the non-overlapping substrings to conclude that $ \myx' $
and $ \myx'' $ are confusable (we illustrate this for the cases when the overlap
happens at the right-hand end of the longer substring, the remaining cases
follow by symmetry):
\begin{inparaenum}
\item[(i)]
for the case of overlapping substrings of lengths $ 1 $ and $ 2 $, write
$ \myx = \bs{u}\,a\,b\,\bs{v} $ and note that its descendants
$ \myx' = \bs{u}\,\overline{a\,b}\,\underline{a\,b}\,\bs{v} $ and
$ \myx'' = \bs{u}\,a\,\overline{b}\,\underline{b}\,\bs{v} $ are
confusable as they have a common descendant $ \bs{u}\,a\,b\,a\,b\,b\,\bs{v} $;
\item[(ii)]
for the case of overlapping substrings of lengths $ 2 $ and $ 2 $, write
$ \myx = \bs{u}\,a\,b\,c\,\bs{v} $ and note that its descendants
$ \myx' = \bs{u}\,\overline{a\,b}\,\underline{a\,b}\,c\,\bs{v} $ and
$ \myx'' = \bs{u}\,a\,\overline{b\,c}\,\underline{b\,c}\,\bs{v} $ are
confusable as they have a common descendant $ \bs{u}\,a\,b\,a\,b\,c\,b\,c\,\bs{v} $;
\item[(iii)]
for the case of overlapping substrings of lengths $ 2 $ and $ 3 $, where
the overlap is of length $ 1 $,
write $ \myx = \bs{u}\,a\,b\,c\,d\,\bs{v} $ and note that its descendants
$ \myx' = \bs{u}\,\overline{a\,b\,c}\,\underline{a\,b\,c}\,d\,\bs{v} $ and
$ \myx'' = \bs{u}\,a\,b\,\overline{c\,d}\,\underline{c\,d}\,\bs{v} $ are
confusable as they have a common descendant $ \bs{u}\,a\,b\,c\,a\,b\,c\,d\,c\,d\,\bs{v} $;
\item[(iv)]
for the case of overlapping substrings of lengths $ 2 $ and $ 3 $, where
the overlap is of length $ 2 $,
write $ \myx = \bs{u}\,a\,b\,c\,\bs{v} $ and note that its descendants
$ \myx' = \bs{u}\,\overline{a\,b\,c}\,\underline{a\,b\,c}\,\bs{v} $ and
$ \myx'' = \bs{u}\,a\,\overline{b\,c}\,\underline{b\,c}\,\bs{v} $ are
confusable as they have a common descendant $ \bs{u}\,a\,b\,c\,a\,b\,c\,b\,c\,\bs{v} $;
\item[(v)]
for the case of overlapping substrings of lengths $ 3 $ and $ 3 $, where
the overlap is of length $ 1 $,
write $ \myx = \bs{u}\,a\,b\,c\,d\,e\,\bs{v} $ and note that its descendants
$ \myx' = \bs{u}\,\overline{a\,b\,c}\,\underline{a\,b\,c}\,d\,e\,\bs{v} $ and
$ \myx'' = \bs{u}\,a\,b\,\overline{c\,d\,e}\,\underline{c\,d\,e}\,\bs{v} $ are
confusable as they have a common descendant $ \bs{u}\,a\,b\,c\,a\,b\,c\,d\,e\,c\,d\,e\,\bs{v} $;
\item[(vi)]
for the case of overlapping substrings of lengths $ 3 $ and $ 3 $, where
the overlap is of length $ 2 $,
write $ \myx = \bs{u}\,a\,b\,c\,d\,\bs{v} $ and note that its descendants
$ \myx' = \bs{u}\,\overline{a\,b\,c}\,\underline{a\,b\,c}\,d\,\bs{v} $ and
$ \myx'' = \bs{u}\,a\,\overline{b\,c\,d}\,\underline{b\,c\,d}\,\bs{v} $ are
confusable as they have a common descendant $ \bs{u}\,a\,b\,c\,a\,b\,c\,d\,b\,c\,d\,\bs{v} $;
\item[(vii)]
for the case of overlapping substrings of lengths $ 1 $ and $ 3 $,
write $ \myx = \bs{u}\,a\,b\,c\,\bs{v} $ and note that its descendants
$ \myx' = \bs{u}\,\overline{a\,b\,c}\,\underline{a\,b\,c}\,\bs{v} $ and
$ \myx'' = \bs{u}\,a\,b\,\overline{c}\,\underline{c}\,\bs{v} $ are
confusable as they have a common descendant $ \bs{u}\,a\,b\,c\,a\,b\,c\,c\,\bs{v} $.
\end{inparaenum}
The only case that was left out from the above list is the case of overlapping
substrings of lengths $ 1 $ and $ 3 $, where the overlap happens in the middle
of the longer substring.
Namely, for $ \myx = \bs{u}\,a\,b\,c\,\bs{v} $, where $ a, b, c \in \A_q $ are
distinct symbols, let
\begin{subequations}
\label{eq:dupltypes}
\begin{alignat}{3}
\label{eq:I}
   &\myx'  &&= \ \bs{u} \, \overline{a \, b \, c} \, \underline{a \, b \, c} \, \bs{v} \\
\label{eq:II}
   &\myx'' &&= \ \bs{u} \, a \, \overline{b} \, \underline{b} \, c \, \bs{v} \ .
\end{alignat}
\end{subequations}
In this case we cannot apply the same reasoning as before to conclude that
$ \myx' $ and $ \myx'' $ are confusable, and indeed they are not in general.
For example, if both $ \bs{u} $ and $ \bs{v} $ are empty strings, then
$ \myx' $ and $ \myx'' $ in \eqref{eq:dupltypes} are non-confusable
because the symbol $ a $ cannot appear after the symbol $ c $ in the descendants
of $ \myx'' $, whereas $ a $ appears after $ c $ in \emph{all} descendants
of $ \myx' $ (a similar example was given in \cite{jain2}).
This situation arises because the segment $ a\,b\,c $ that appears in the
original string $ \myx $ no longer appears in $ \myx'' $ as it has been
``broken up'' by the insertion of a copy of $ b $.
In conclusion, one can mimic in $ \myx'' $ (resp.\ $ \myx'$) the duplication
that has produced $ \myx' $ (resp.\ $ \myx'' $) from $ \myx $, and thus conclude
that $ \myx' $ and $ \myx'' $ are confusable, in all cases with the exception
of \eqref{eq:dupltypes}.
Therefore, a zero-error code in $ D^1(\myx) $ can contain at most $ 2 $ codewords.
\end{proof}

\begin{remark}
\label{rem:dupltypes}
\textnormal{
Not every situation of the form \eqref{eq:dupltypes} will result in
non-confusable descendants.
As a counterexample, suppose that $ \bs{u} $ is empty and $ \bs{v} = a $,
so that $ \myx = a\,b\,c\,a $, $ \myx' = \overline{a\,b\,c}\,\underline{a\,b\,c}\,a $,
and $ \myx'' = a\,\overline{b}\,\underline{b}\,c\,a $.
Now $ \myx' $ and $ \myx'' $ have a common descendant
$ a\,b\,b\,c\,a\,b\,c\,a $ and are therefore confusable.
However, the fact that \eqref{eq:dupltypes} is \emph{the only} case when
two descendants \emph{may} be non-confusable is sufficient for our purposes.
In particular, it will ena\-ble us to derive a tight \emph{upper} bound on
the cardinality of optimal zero-error codes.
\myqed
}
\end{remark}

The above observation is true in general, not just for $ 1 $-descendants
of a string $ \myx $.
Namely, if $ \myx', \myx'' \in D^*(\myx) $ are obtained by applying two
different patterns of duplications on $ \myx $, in each of these strings
we can repeat/imitate the duplications applied to the other, in the same
order, and thus conclude that they have a common descendant.
The only way this imitation process \emph{may} become impossible from some
point on, is to arrive at a situation where a duplication of length $ 3 $
has been applied in $ \myx' $ and a duplication of length $ 1 $ on the
appropriate segment in $ \myx'' $ (see \eqref{eq:dupltypes}), so that
$ \myx'' $ is not able to imitate the corresponding mutation in $ \myx' $.
This is because, whenever a duplication of length $ 2 $ or $ 3 $ is performed
in a string, \emph{all} segments of length $ \leqthree $ from the original
string are preserved in the resulting string (with additional few substrings
being created at the place the duplicate was inserted).
The only way for a segment of length $ 3 $ from the original string
to disappear in the resulting string is after a duplication of length $ 1 $,
as illustrated in \eqref{eq:II}.
Based on this observation, we shall derive an upper bound on the cardinality
of optimal zero-error codes in the set of all $ t $-descendants of a given
string $ \myx $, for any $ t $ (Proposition~\ref{thm:desc} ahead).

\begin{example}
Here is the example from \eqref{eq:desc0} presented in a slightly different
way so as to further clarify our point
(the segment $ {\color{blue} 1 \, 2 \, 0} $ is highlighted, and the duplications
to the left, resp. right, of this segment are shown so that a duplicate is
inserted to the left, resp. right, of the original):
\begin{subequations}
\label{eq:desc}
\begin{alignat}{3}
\label{eq:desc1}
  &\myx            &&=  \ \phantom{0 \ 0 \ 0} \ 0 \ &&1 \ {\color{blue} 1 \ 2 \ 0} \ 2 \ 1 \ 0   \\
  &\myx^\sml{(1)}  &&=  \ \phantom{0 \ 0} \ \underline{0} \ \overline{0} \ &&1 \ {\color{blue} 1 \ 2 \ 0} \ 2 \ 1 \ 0   \\
\label{eq:x2}
	&\myx^\sml{(2)}  &&=  \ \phantom{0 \ 0} \ 0 \ 0 \ &&1 \ {\color{blue} 1} \ \overline{ {\color{blue} 2 \ 0} \ 2 } \ \underline{ 2 \ 0 \ 2 } \ 1 \ 0  \\
	&\myx^\sml{(3)}  &&=  \ \phantom{0 \ 0} \ 0 \ 0 \ &&1 \ {\color{blue} 1 \ 2 \ 0} \ 2 \ 2 \ 0 \ \overline{ 2 \ 1} \ \underline{ 2 \ 1 } \ 0  \\
  &\myx^\sml{(4)}  &&=  \ 0 \ 0 \ \underline{ 1 \ 1 } \ &&\overline{ 1 \ {\color{blue} 1} } \ {\color{blue} 2 \ 0} \ 2 \ 2 \ 0 \ 2 \ 1 \ 2 \ 1 \ 0  \ .
\end{alignat}
\end{subequations}
Let
$ \myz \, = \ 0 \ 1 \ {\color{blue} 1 \ 2} \ {\color{red} 2} \ {\color{blue} 0} \ 2 \ 1 \ 0 $.
Note that $ \myz $ can mimic all duplications of substrings of $ \myx $
that either do not overlap with the segment $ {\color{blue} 1 \, 2 \, 0} $,
or overlap with it only partially%
\footnote{For $ \myx $ in \eqref{eq:desc1}, the substrings of length
$ \leqthree $ that partially overlap with the substring $ {\color{blue} 1 \, 2 \, 0} $
are: $ {\color{blue} 1} $, $ {\color{blue} 2} $, $ {\color{blue} 0} $,
$ 1 \,{\color{blue} 1} $, $ {\color{blue} 1 \, 2} $, $ {\color{blue} 2 \, 0} $,
$ {\color{blue} 0} \, 2 $, $ 0 \, 1 \, {\color{blue} 1} $,
$ 1 \, {\color{blue} 1 \, 2} $, $ {\color{blue} 2 \, 0} \, 2 $,
$ {\color{blue} 0} \, 2 \, 1 $.
},
such as those illustrated in \eqref{eq:desc}:
\begin{subequations}
\label{eq:desc2}
\begin{alignat}{3}
  &\myx             &&=  \ \phantom{0 \ 0 \ 0} \ 0 \ &&1 \ {\color{blue} 1 \ 2 \ 0} \ 2 \ 1 \ 0   \\
  &\myz             &&=  \ \phantom{0 \ 0 \ 0} \ 0 \ &&1 \ {\color{blue} 1} \ \overline{{\color{blue} 2}} \ \underline{{\color{red} 2}} \ {\color{blue} 0} \ 2 \ 1 \ 0   \\
  &\myz^\sml{(1)}   &&=  \ \phantom{0 \ 0} \ \underline{0} \ \overline{0} \ &&1 \ {\color{blue} 1 \ 2} \ {\color{red} 2} \ {\color{blue} 0} \ 2 \ 1 \ 0   \\
	&\myz^\sml{(2)}   &&=  \ \phantom{0 \ 0} \ 0 \ 0 \ &&1 \ {\color{blue} 1 \ 2} \ \overline{ {\color{red} 2} \ {\color{blue} 0} \ 2 } \ \underline{ 2 \ 0 \ 2 } \ 1 \ 0  \\
	&\myz^\sml{(3)}   &&=  \ \phantom{0 \ 0} \ 0 \ 0 \ &&1 \ {\color{blue} 1 \ 2} \ {\color{red} 2} \ {\color{blue} 0} \ 2 \ 2 \ 0 \ \overline{ 2 \ 1} \ \underline{ 2 \ 1 } \ 0  \\
  &\myz^\sml{(4)}   &&=  \ 0 \ 0 \ \underline{ 1 \ 1 } \ &&\overline{ 1 \ {\color{blue} 1} } \ {\color{blue} 2} \ {\color{red} 2} \ {\color{blue} 0} \ 2 \ 2 \ 0 \ 2 \ 1 \ 2 \ 1 \ 0  \ .
\end{alignat}
\end{subequations}
Therefore, any pair of strings from \eqref{eq:desc} and \eqref{eq:desc2} are
confusable;
for example, a common descendant of $ \myz $ and $ \myx^\sml{(3)} $ above is
$ \myz^\sml{(3)} $.
The only mutation $ \myz $ cannot imitate is the duplication of the entire
segment $ {\color{blue} 1\,2\,0} $ because the corresponding segment
in $ \myz $ no longer exists (it has been ``broken up'' by the inserted
symbol $ {\color{red} 2} $).
For example, if $ \myx^\sml{(2)} $ in \eqref{eq:x2} was to mutate to
\begin{equation}
  \myy \ = 0 \ 0 \ 1 \ \overline{ {\color{blue} 1 \ 2 \ 0} } \ \underline{ 1 \ 2 \ 0 } \ 2 \ 2 \ 0 \ 2 \ 1 \ 0
\end{equation}
instead of $ \myx^\sml{(3)} $, it would no longer be possible to apply the
same process as in \eqref{eq:desc2}.
\myqed
\end{example}

Before stating Proposition~\ref{thm:desc}, which is the main result of this
section, we prove a useful lemma.

\begin{lemma}
\label{thm:seq}
Fix positive integers $ b, t, n $ with $ b \leqslant t \leqslant n $.
Let $ {\mathcal U} \subseteq \{ l_1, l_3, *\}^n $ be a set of strings
satisfying the following two conditions:
\begin{inparaenum}
\item[1)]
every string $ \bs{u} \in {\mathcal U} $ has exactly $ b $ symbols $ l_3 $,
$ t-b $ symbols $ l_1 $, and $ n - t $ symbols $ * $, and
\item[2)]
for every two distinct strings $ \bs{u}, \bs{v} \in {\mathcal U} $ there is a
location $ i \in \{1, \ldots, n\} $ at which $ \{u_i, v_i\} = \{l_1, l_3\} $
(i.e., such that either $ u_i = l_1, v_i = l_3 $, or $ u_i = l_3, v_i = l_1 $).
\end{inparaenum}
Then $ |{\mathcal U}| \leqslant 2^{t H(b/t)} $.
\end{lemma}
\begin{proof}
Consider $ n $ tosses of a coin whose probability of landing heads is $ b/t $,
and define the following events indexed by the strings in $ {\mathcal U} $.
For $ \bs{u} \in {\mathcal U} $, let $ A_{\bs{u}} $ be the event of the $ i $'th
toss landing heads if $ u_i = l_3 $, landing tails if $ u_i = l_1 $, and landing
arbitrarily if $ u_i = * $, for all $ i = 1, \ldots, n $.
It follows from the condition 1) that the probability of this event is
$ \Prob\{A_{\bs{u}}\} = \big( \frac{b}{t} \big)^b \big( 1 - \frac{b}{t} \big)^{t-b} = 2^{- t H(b/t)} $
for every $ \bs{u} \in {\mathcal U} $.
Furthermore, by the condition 2), the events $ A_{\bs{u}} $ and $ A_{\bs{v}} $
are disjoint for any $ \bs{u}, \bs{v} \in {\mathcal U} $, $ \bs{u} \neq \bs{v} $.
Therefore, $ |{\mathcal U}| \cdot \Prob\{A_{\bs{u}}\} \leqslant 1 $, which is
what we wanted to prove.
\end{proof}

Note that the stated upper bound on $ |{\mathcal U}| $ does not depend on $ n $
(i.e., on the number of $ * $ symbols).

\begin{proposition}
\label{thm:desc}
Consider a string $ \myx \in \A_q^n $, and let $ \C \subseteq D^t(\myx) $
be a zero-error code for the $ (\leqthree) $-tandem-duplication channel
satisfying the requirement that, out of $ t $ duplications producing each
codeword $ \myy \in \C $ from $ \myx $, exactly $ b $ are of length $ 3 $.
Then $ |\C| \leqslant 2^{t H(b/t)} $.
\end{proposition}
\begin{proof}
As shown above, if two strings $ \myx', \myx'' \in D^t(\myx) $ are
non-confusable, then it necessarily holds that a duplication of length $ 3 $
was applied on a segment $ a \, b \, c $ in one of the ancestors of $ \myx' $,
while a duplication of length $ 1 $ was applied on the middle symbol of the
corresponding segment in an ancestor of $ \myx'' $, or vice versa.
In other words, for every pair of codewords of a zero-error code in $ D^t(\myx) $
there is a segment on which they differ by a length~$ 1 $/length~$ 3 $ mutation.
Hence, the question we are interested in is how big a set of strings, any two of which
differ by a length 1/length 3 mutation at some location, can be.
(It is irrelevant which locations are these, and what happens in between
them, the only requirement we have is that each pair of codewords differs
somewhere by a length 1/length 3 mutation because this is the only way two
strings may become non-confusable.)
Therefore, $ |\C| $ can be upper bounded by the maximum cardinality of a set
$ {\mathcal U} $ of strings over the ``alphabet''
$ \{\textnormal{length $ 1 $}, \textnormal{length $ 3 $}, *\} $
satisfying the following conditions:
\begin{inparaenum}
\item[1)]
every string in $ {\mathcal U} $ has exactly $ b $ symbols `length $ 3 $' and
$ t - b $ symbols `length $ 1 $', and
\item[2)]
every two distinct strings in $ {\mathcal U} $ differ at some location by
an `length~$ 1 $'/`length~$ 3 $' symbol.
\end{inparaenum}
(`$ * $' is a dummy symbol that serves to fill in the empty locations which
may occur because different pairs of codewords may differ by a length~$ 1 $/length~$ 3 $
mutation on different segments; see also Example~\ref{ex:l1l3} below.)
The stated upper bound on $ |\C| $ is now obtained by invoking Lemma~\ref{thm:seq}.
\end{proof}

\begin{example}
\label{ex:l1l3}
Let us illustrate the set of strings $ \mathcal{U} $ mentioned in the preceding
proof.
Consider the following string:
\begin{align}
\label{eq:xseg}
  \myx\phantom{^\sml{(1)}}  =  \ \rlap{$\aoverbrace[L1R]{\phantom{0 \ 1 \ 2}}$} 0 \, \aunderbrace[l1r]{1 \ 2 \ 3} \, 4 \  \rlap{$\aoverbrace[L1R]{\phantom{5 \ 6 \ 7}}$} 5 \ 6 \, \aunderbrace[l1r]{7 \ 8 \ 9}  \ . \phantom{6 \ 6 \ 7 \ 8 \ 9}
\end{align}
(All the symbols of $ \myx $ being different makes the example clearer without
affecting its generality.)
Let the following descendants in $ D^3(\myx) $ be obtained by applying tandem duplications on
the four segments in $ \myx $ that are under- or over-braced:
\begin{subequations}
\label{eq:desc3}
\begin{alignat}{2}
  &\myx'    &&=  \ 0 \ 1 \ 2 \ \underline{0 \ 1 \ 2} \ \underline{2} \ 3 \ 4 \ 5 \ 6 \ \underline{6} \ 7 \ 8 \ 9  \\
	&\myx''   &&=  \ 0 \ 1 \ \underline{1} \ 2 \ 3 \ 4 \ 5 \ 6 \ \underline{6} \ 7 \ 8 \ 9 \ \underline{7 \ 8 \ 9}  \\
	&\myx'''  &&=  \ 0 \ 1 \ \underline{1} \ 2 \ 3 \ 4 \ 5 \ 6 \ 7 \ \underline{5 \ 6 \ 7} \ 8 \ \underline{8} \ 9 \ ,
\end{alignat}
\end{subequations}
where the inserted duplicates are underlined.
The mutations applied on the four segments can be described by using the
following strings:
\begin{subequations}
\label{eq:desc4}
\begin{alignat}{6}
  &\bs{u}'    &&=  \ \ &&l_3 \ &&l_1 \ &&l_1 \ &&*  \\
	&\bs{u}''   &&=  \ \ &&l_1 \ &&* \ &&l_1 \ &&l_3  \\
	&\bs{u}'''  &&=  \ \ &&l_1 \ &&* \ &&l_3 \ &&l_1 \ ,
\end{alignat}
\end{subequations}
where the symbol $ l_3 $ indicates that a duplication of length $ 3 $ was
applied on the observed segment, $ l_1 $ indicates that a duplication of
length $ 1 $ was applied on the middle symbol of that segment, and $ * $
indicates that neither of those two mutations was applied on that segment.
Note that, for every pair of strings in \eqref{eq:desc4}, there is a coordinate
at which one of them equals $ l_1 $ while the other equals $ l_3 $.
\myqed
\end{example}

\section{Zero-error capacity of the $ (\leqthree) $-tandem-duplication channel}

Let $ \C^\star_{q}(n) \subseteq \A_q^n $ be an optimal zero-error code for the
$ (\leqthree) $-tandem-duplication channel.
For a given irreducible string $ \myx \in \Irr_q $, define
$ \C^\star_{q}(n; \myx) \defeq \C^\star_{q}(n) \cap D^*(\myx) $.
Then $ \C^\star_{q}(n; \myx) $ is an optimal zero-error code in the set of all
descendants of $ \myx $ of length $ n $.
This is because, due to the root-uniqueness property for tandem duplications
of length $ \leqthree $, any two different descendant cones are disjoint
\cite[Cor.~26]{jain2}, and hence any two strings having different roots are
non-confusable.
In other words, one can, without loss of generality, construct a code separately
in the descendant cones of each of the possible roots/irreducible strings.
This fact is stated explicitly in the following lemma; the proof is omitted
as it follows directly from \cite[Cor.~26]{jain2}.

\begin{lemma}
\label{thm:x}
An optimal zero-error code of length $ n $ can be expressed as a disjoint
union of optimal codes in each of the descendant cones:
\begin{align}
\label{eq:Cstar}
  \C^\star_{q}(n)  =  \dot\bigcup_{\myx \in \Irr_q} \C^\star_{q}(n; \myx) .
\end{align}
\end{lemma}

The following claim gives a characterization of the exponential growth-rate
of the cardinality of optimal codes $ \C^\star_{q}(n) $ or, equivalently, of
the zero-error capacity of the $ (\leqthree) $-tandem-duplication channel.
It states that this quantity equals
$ \iota_q = \lim_{n \to \infty} \frac{1}{n} \log_2 I_q(n) $
(see \eqref{eq:expirr} and \eqref{eq:r}).
In other words, the zero-error capacity is attained by the codes $ \Irr_q(n) $
consisting of irreducible strings of length $ n $.

\begin{theorem}
\label{thm:main}
The zero-error capacity of the $ (\leqthree) $-tandem-duplication channel
with alphabet $ \A_q $, $ q \geqslant 3 $, equals $ \iota_q $.
\end{theorem}
\begin{proof}
We need to show that:
\begin{equation}
\label{eq:rate}
  \lim_{n \to \infty} \frac{1}{n} \log_2\!\left| \C^\star_{q}(n) \right|
    =  \iota_q .
\end{equation}
Since $ \Irr_q(n) \subseteq \C_q^\star(n) $, we know that
$ \lim_{n \to \infty} \frac{1}{n} \log_2 | \C_q^\star(n) | \geqslant
  \lim_{n \to \infty} \frac{1}{n} \log_2 I_q(n)  =  \iota_q $
(see \eqref{eq:expirr}), so it is enough to prove the opposite inequality
$ \lim_{n \to \infty} \frac{1}{n} \log_2 | \C_q^\star(n) | \leqslant \iota_q $.
In order to show this, we shall simplify the analysis by constructing a
sufficiently large \emph{subcode} $ \C_q(n; m,t,b) \subseteq \C_q^\star(n) $
having the same exponential growth-rate as the optimal code $ \C_q^\star(n) $,
i.e.,
$ \lim_{n \to \infty} \frac{1}{n} \log_2 | \C^\star_{q}(n) | =
  \lim_{n \to \infty} \frac{1}{n} \log_2 | \C_q(n; m,t,b) | $,
for an appropriate choice of the parameters $ m, t, b $.

Fix an arbitrary irreducible string $ \myx $ of length $ m $, $ \myx \in \Irr_q(m)$,
and let $ \C_{q}(n; \myx, t,b) \subseteq \C^\star_{q}(n; \myx) $ be a code
containing only those codewords of $ \C^\star_{q}(n; \myx) $ that satisfy the
following two conditions:
\begin{inparaenum}
\item[1)]
every codeword belongs to $ D^t(\myx) $, i.e., is a $ t $-descendant of
$ \myx $, and
\item[2)]
out of $ t $ duplications producing a given descendant/codeword from $ \myx $,
exactly $ b $ are of length~$ 3 $.
\end{inparaenum}
We then define the above-mentioned subcode as:
\begin{align}
\label{eq:C}
  \C_q(n; m,t,b)  \defeq  \bigcup_{\myx \in \Irr_q(m)} \C_{q}(n; \myx, t,b) .
\end{align}
It follows from the construction and Lemma~\ref{thm:x} that:
\begin{align}
\label{eq:C1}
  \C_q^\star(n) = \bigcup_{m,t,b} \C_q(n; m,t,b) .
\end{align}
It should now be clear that $ | \C_q(n; m,t,b) | $, maximized over
all possible values of $ m, t, b $, has the same exponential growth-rate
as $ | \C^\star_{q}(n) | $ (the choice of $ m, t, b $ is made
for every $ n $, i.e., the optimal values of the parameters $ m, t, b $
are in general functions of the block-length $ n $).
This follows from \eqref{eq:C1} and the pigeon-hole principle---the
cardinality of the code $ \C^\star_{q}(n) $ grows exponentially fast in the
block-length $ n $, and there are linearly many choices for each of $ m $,
$ t $, and $ b $, so for at least one of these choices the codes $ \C_q(n; m,t,b) $
will contain exponentially many codewords (with the same exponent).
Therefore, the codes $ \C_q(n; m,t,b) $ are asymptotically optimal in terms
of rate, i.e., they achieve the zero-error capacity of the
$ (\leqthree) $-tandem-duplication channel, when the parameters $ m, t, b $
are chosen appropriately (so as to maximize $ | \C_q(n; m,t,b) | $).

Let us now calculate the rate of the constructed codes.
By \eqref{eq:C} and Proposition~\ref{thm:desc} (which states that
$ | \C_{q}(n; \myx, t,b) | \leqslant 2^{t H(b/t)} $), the cardinality
of the code $ \C_q(n; m,t,b) $ can be upper-bounded as:
\begin{align}
\label{eq:Csize}
  \big| \C_q(n; m,t,b) \big| \leqslant I_q(m) \cdot 2^{t H(b/t)} ,
\end{align}
while the length of this code can be lower-bounded as:
\begin{align}
\label{eq:length}
  n  \geqslant  m + 3b + (t-b)  =  m + t + 2b
\end{align}
(the initial irreducible string is of length $ m $, and exactly $ b $
duplications that produce its descendants are of length $ 3 $).
Therefore,
\begin{align}
\label{eq:Crate}
  \frac{1}{n} \log_2\!\big| \C_q(n; m,t,b) \big|  \leqslant
	 \frac{ \log_2 I_q(m) + t H(b/t) }{ m + t + 2b } .
\end{align}
To determine the asymptotics of this quantity as $ n \to \infty $,
two cases that correspond to different choices of the parameters
$ m, t, b $ need to be considered:
\begin{itemize}[leftmargin=0.6cm]
\item
$ m = o(t) $.
Let $ \lim_{t \to \infty} \frac{b}{t} = \beta \in [0,1] $.
Then:
\begin{align}
  \limsup_{n \to \infty} \frac{1}{n} \log_2 \!\big| \C_q(n; m,t,b) \big|
    \leqslant  \frac{H(\beta)}{1 + 2\beta}
    \leqslant  \iota_q ,
\end{align}
where the first inequality follows from \eqref{eq:Crate}, and the second
is identical to \eqref{eq:entropy}.
\item
$ t = {\mathcal O}(m) $.
Let $ \liminf_{m \to \infty} \frac{t}{m} = \tau \geqslant 0 $ and
$ \lim_{t \to \infty} \frac{b}{t} = \beta \in [0,1] $.
Then:
\begin{align}
  \limsup_{n \to \infty} \frac{1}{n} \log_2 \!\big| \C_q(n; m,t,b) \big|
    \leqslant  \frac{ \iota_q + \tau H(\beta) }{ 1 + \tau(1 + 2\beta) }
    \leqslant  \iota_q .
\end{align}
Again, the first inequality follows from \eqref{eq:Crate}, and the second
is equivalent to \eqref{eq:entropy}.
\end{itemize}
In conclusion, all choices of the parameters $ m, t, b $ result in the
asymptotic rate of the codes $ \C_q(n; m,t,b) $ being $ \leqslant\!\iota_q $.
Since these codes are rate-wise optimal, as argued
in the second paragraph of this proof, the identity \eqref{eq:rate} is
thereby established.
\end{proof}

\pagebreak
\section{Conclusion}

The evolution of strings under tandem duplications is an interesting and
non-trivial problem of relevance in several fields of research.
In this work, we have studied the confusability of strings under tandem
duplications of varying length, a problem inspired by error correction in
communication channels in which the transmitted messages are affected by
this kind of mutations.
Specifically, for the case of duplications of length $ \leqthree $, we have
derived an upper bound on the maximum cardinality of a set of pairwise
non-confusable strings, which, together with the construction from \cite{jain2},
establishes the maximum rate achievable by codes correcting an \emph{arbitrary}
number of such impairments.

In cases in which duplication roots are not unique, e.g., the
$ (\leqslant\!\ell) $-tandem-duplication models with parameter $ \ell $
larger than $ 3 $, the maximum achievable rates remain unknown.
Due to the ``root non-uniqueness property'', the analysis of the evolution
and confusability of strings in these models is more complicated and,
hence, further work, and possibly different methods, will be required
to solve the zero-error capacity and related problems therein.

\vspace{5mm}
\bibliographystyle{amsplain}

\end{document}